\documentclass{article}

\usepackage{arxiv}

\usepackage[utf8]{inputenc} 
\usepackage[T1]{fontenc}    
\usepackage{hyperref}       
\usepackage{url}            
\usepackage{booktabs}       
\usepackage{amsfonts}       
\usepackage{amsmath}       
\usepackage{nicefrac}       
\usepackage{microtype}      
\usepackage{lipsum}
\usepackage{graphicx}
\usepackage{calc}
\usepackage{xcolor}

\graphicspath{ {./images/} }

\title{Implementing Simulation of Simplicity for geometric degeneracies}

\author{W. Randolph Franklin\\
Electrical, Computer, and Systems Engineering Dept.\\
Rensselaer Polytechnic Institute\\
Troy, NY 12180\\
\texttt{mail@wrfranklin.org}\\
\And
Salles Viana Gomes de Magalhães\\
Departamento de Informática\\
Universidade Federal de Viçosa\\
Viçosa MG Brazil\\
\texttt{sallesviana@gmail.com}
}

\DeclareGraphicsExtensions{.png,.pdf}
\graphicspath{ {images/} }
\usepackage{wrapfig}

\begin{document}
\maketitle

\begin{abstract}
  We describe how to implement Simulation of Simplicity (SoS).  SoS removes geometric degeneracies in point-in-polygon queries, polyhedron intersection, map overlay, and other 2D and 3D geometric and spatial algorithms by determining the effect of adding non-Archimedian infinitesimals of different orders to the coordinates.  Then it modifies the geometric predicates to emulate that, and evaluates them in the usual arithmetic.

  A geometric degeneracy is a coincidence, such as a vertex of one polygon on an edge of another polygon, that would have probability approaching zero if the objects were distributed i.i.d. uniformly.  However, in real data, they can occur often.  Especially in 3D, there are too many types of degeneracies to reliably enumerate.  But, if they are not handled, then predicates evaluate wrong, and the output topology may be wrong.

We describe the theory of SoS, and how several algorithms and  programs were successfully modified, including volume of the union of many cubes, point location in a 3D mesh, and intersecting 3D meshes.
\end{abstract}


\section{Introduction}
Simulation of Simplicity (SoS) removes geometric degeneracies (special cases) by determining the effect of adding non-Archimedian infinitesimals of different orders to coordinates, and then modifying geometric predicates to simulate that.  The modified predicates will then be evaluated in the usual arithmetic.  In other words, the effect of using infinitesimals on the value of a predicate is determined and then the predicate is recoded to compute, in the usual arithmetic, what its new result would be if the original predicate were executed using infinitesimals.

\begin{wrapfigure}[11]{r}{.2\textwidth}
\vspace{-1em}
\includegraphics[width=.2\textwidth]{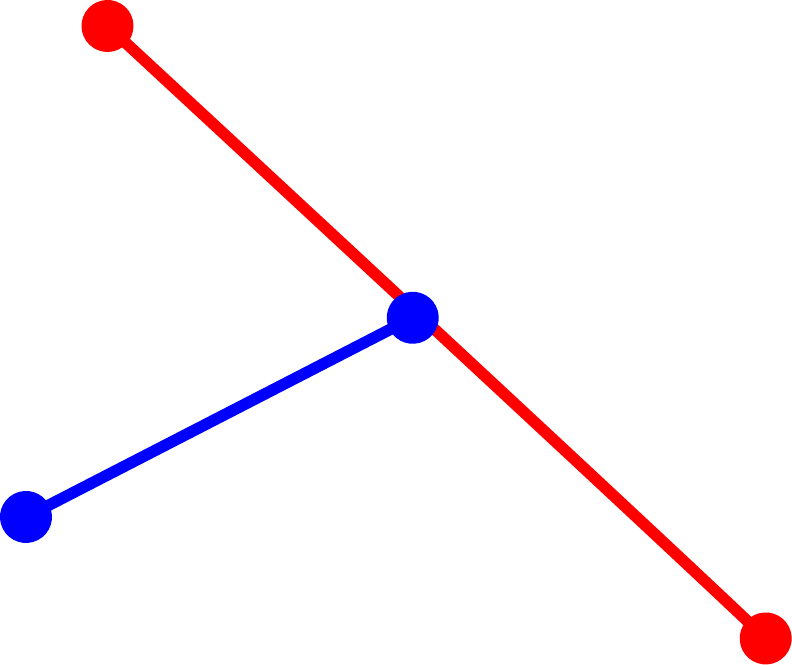}
  \caption{\small Degeneracy: endpoint of blue edge on red edge}
  \label{f:degen}
\end{wrapfigure}
Handling geometric special cases is a nasty part of transforming a beautiful algorithm into useful code.  Example 2D degeneracies include an endpoint of one edge on another edge (Figure \ref{f:degen}),
a query point directly below a poluhedron vertex (see 
Section \ref{s:pnpoly} on Page \pageref{s:pnpoly}),
two overlapping edges, two coincident points from different objects, two edges from different objects with a common vertex, etc.  If objects were randomly uniformly and independently and identically distributed, then the probability of any of these would be $0$.  However they occur frequently in real data.  When a CAD designer place one object tight against another, a geometric degeneracy is created.  Two diplomats may agree that the common border between their adjacent countries will coincide with a river shoreline or center line.  Two different digital maps created for the same region may have many (but not all) edges in common.  We may want to conflate a dataset of contour lines with a dataset of hydrography features, including shorelines, where the shoreline might mostly align with a contour line.

The approach described in this note is better than existing solutions because not handling degeneracies correctly causes erroneous output (examples are shown in some of the references), and allegedly, even can cause commercial CAD systems to crash.  One quick fix is to perturb coordinates by a small amount.
Heuristics like this work only up to a point, and fail for more complicated algorithms.  Indeed, then there are too many special cases to understand.  Also, as input datasets get larger, the probability of an error grows.

\begin{wrapfigure}[13]{r}{.2\textwidth}
\includegraphics[width=.2\textwidth]{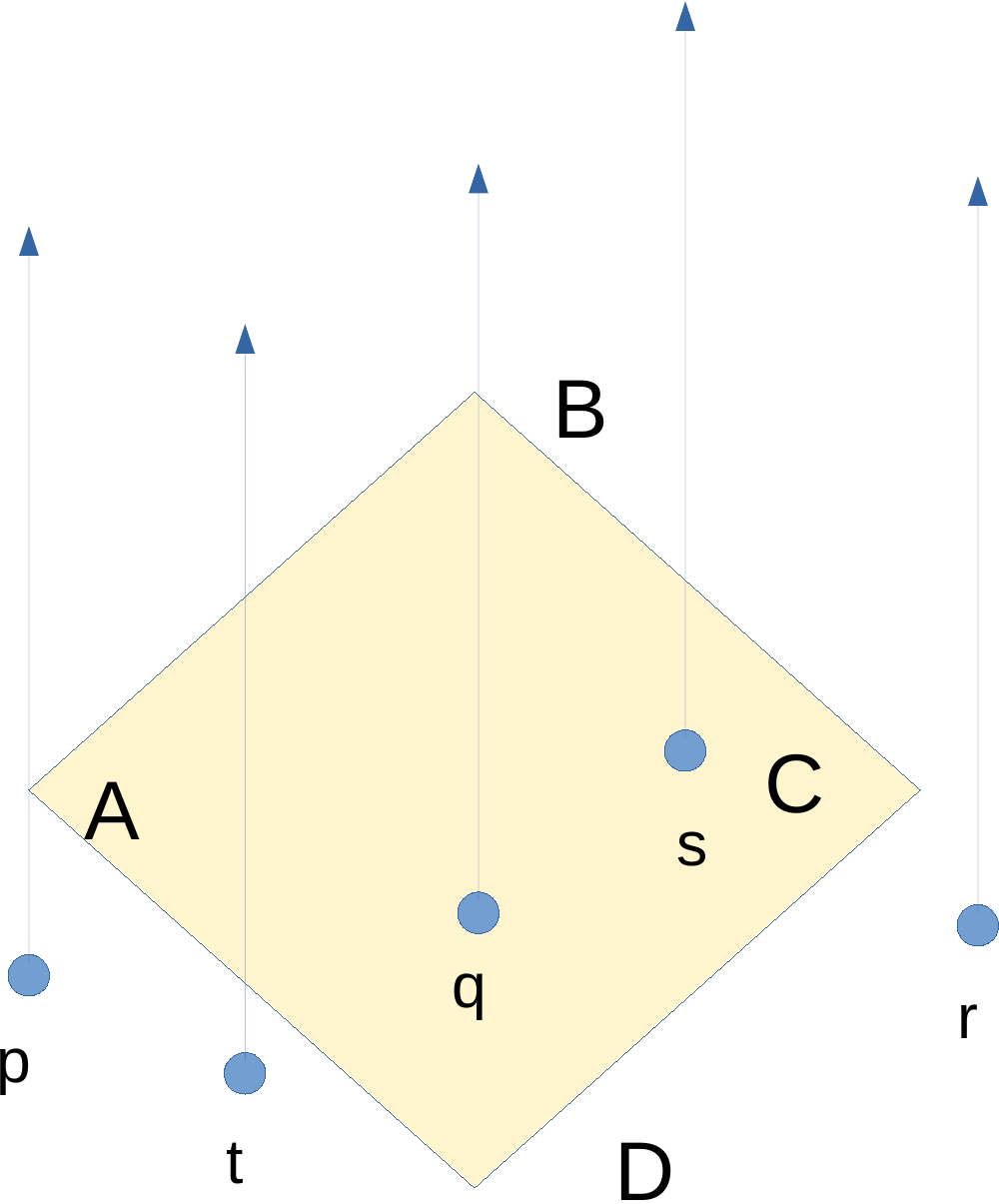}
  \caption{\small Several cases for testing point in polygon}
  \label{f:pnpoly}
\end{wrapfigure}

Why is this difficult?  Consider the simple 2D problem of determining whether a point $p$ is contained inside a polygon $\mathcal P$ (Figure \ref{f:pnpoly}), \cite{pnpoly-url}, also called \emph{ray casting}.  Here we introduce the problem; how to solve it with SoS is presented in Section \ref{s:pnpoly} on Page \pageref{s:pnpoly}.
The classic Jordan-curve algorithm\cite{jordan1887}, extends a semi-infinite ray  up from $p$, and counts how many edges $\mathbf{e}_i$ of $\mathcal P$ it crosses.  Point $p$ is inside $\mathcal P$ iff that number is odd.  (We will not here consider what is a legal polygon, and what is the answer if $p$ is on an edge.)

But, what if  the ray intersects an endpoint of  $\mathbf{e}_i$?  Is that an intersection or not?   We may not care about one individual  $\mathbf{e}_i$.  We may not even care about the exact total number of intersections.  However, we require that the total be odd or even correctly.   This requires that for some rays and  $\mathbf{e}_i$, there must be an intersection, and for others there must not.    One heuristic that seems to work to count an intersection iff the ray intersects the right endpoint of $\mathbf{e}_i$.   If  $\mathbf{e}_i$ is vertical, then neither endpoint intersects the ray.



This will give the wrong answer when $p$ is close to an edge of $\mathcal P$ and $p'$ is on the other side.   However the probability of that error is reduced by reducing how much $p$ was translated.   Importantly, the ray run up from $p'$ will never intersect any $\mathbf{e}_i$, so our difficult  degeneracy will never occur.

The above analysis illustrates the divide-and-conquer technique.  We decomposed the problem of point location in a polygon into a set of smaller problems of testing the point (actually, its ray) separately against each edge.   Then we combined the answers to the smaller problems to solve the original problem.

Locating a point in a polygon is so simple that the time spent above on analyzing the special case may seem excessive.  We (think that we) can easily make the algorithm work.  However, the problem of using the Jordan curve algorithm to test whether a point $p$ is contained in a 3D polyhedron $\mathcal P$ has enough special cases that confidently handling them all is problematic.  The general algorithm is to run a ray $r$ up from $p$ and test which faces it intersects.  $p$ is inside iff the number of intersections is odd.  Some special cases are as follows.  (i) $p$ might be on an edge, or on a vertex of $\mathcal P$. (ii) $r$ might intersect an edge, in any of several different ways. (iii) $r$ might coincide with an edge, i.e., run along the edge. (iv) $r$ might intersect a vertex.

\begin{wrapfigure}[9]{r}{.2\textwidth}
\includegraphics[width=.2\textwidth]{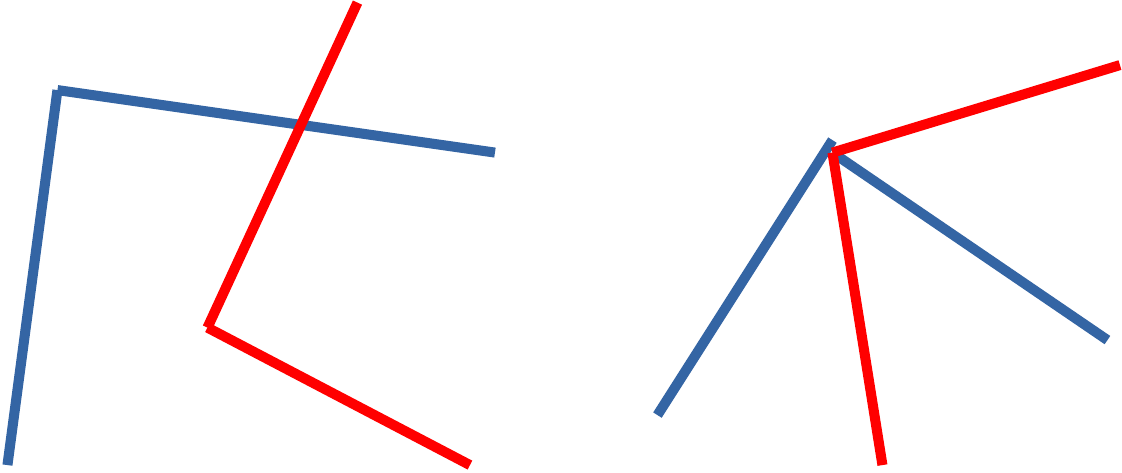}
  \caption{\small A red polyline crossing a blue polyline: the general case, and a degeneracy}
  \label{f:polylines}
\end{wrapfigure}

Another example with enough special cases that an informal analysis is likely to be incomplete and so wrong is the problem of intersecting two polylines $\mathbf{l_0}$ and $\mathbf{l_1}$; see Figure \ref{f:polylines}.  Polyline $\mathbf{l_i}$ is a sequence of $n_i$ vertices $v_{ij}$, $0\le j<n_i$, each defining $n_i-1$ edges $v_{ij}v_{i,j+1}$ for $0\le j<n_i-1$.  The polyline is closed when the last vertex equals the first: $v_{i0} = v_{i,n-1}$.

Assume that each polyline has no two vertices in common; $i\ne k \implies v_{ij}\ne v_{ik}$.  In general position, $\mathbf{l_0}$ has no vertices in common with $\mathbf{l_1}$ and no vertex of either polyline is in the interior of any edge of the other.  That is, if an edge of $\mathbf{l_0}$ intersects (in a point set sense) an edge of $\mathbf{l_1}$, then the intersection is exactly one point.

If a vertex of one polyline were on an edge of the other, that would be a degeneracy; see Figure \ref{f:polylines}.  One application of polyline intersection would be computing a boolean combination of two polygons, where their boundaries are closed polylines.

Another way of looking at a degeneracy is that it occurs when evaluating a boolean predicate that compares two numbers, e.g., is $a<b$?   In this note, assume computations are exact; we do not consider roundoff errors.    This is not to say that roundoff errors are unimportant, merely that they are another topic, which we also treat in some our cited applications.   SoS assumes exact computation, which can be achieved, e.g., by computing with big rational numbers.   Rational numbers represent each number as the ratio of two integers, e.g., $1/3$, and compute exactly, e.g., $1/3+2/5=11/15$.

Geometric predicate evaluations can be considered as determinant sign evaluations.   Consider three 2D points, $A, B, C$.   One possible predicate is whether $\angle ABC < \pi$?   That, with the other information, controls whether the convex hull of $OABC$ is $OABC$ or $OAC$.   $\angle ABC < \pi$ is equivalent to \scalebox{.7}{$\begin{vmatrix} A_x & A_y & 1 \\ B_x & B_y & 1 \\ C_x & C_y & 1 \end{vmatrix} > 0$} .

The problem of evaluating whether $a<b$ is that $a=b$ is a third case\footnote{The notations for equality and assignment are not standardized.  We will use $=$ for the equality predicate, and $\leftarrow$ for assignment.}.   Every decision tree that branched out two ways at each decision, now has to branch three ways to handle the degeneracies.    After $k$ decisions, $2^k$ cases grows to $3^k$ cases.   One possible solution is to fold the degenerate case $a=b$ into one of the two nondegenerate cases.  That idea is an informal step towards SoS.

In brief, the solution presented in this note is a technique for modifying a geometry algorithm (and so also its source code) so that it will handle degenerate geometric inputs with effectively no increase in execution time.

\section{Infinitesimals}

Here we extend the set of real numbers $\Re$ into a \emph{non-Archimedian ordered field}\cite{enwiki:972176701} by adding \emph{infinitesimals}.

$\Re$ may be partitioned into negative finite numbers, zero, and positive finite numbers.  A positive \emph{infinitesimal}, $\epsilon$, is smaller than any positive real number.  That is $\forall r\in\Re, r>0\implies0<\epsilon<r$.  This is logically possible.  One way to approach this is to consider $\epsilon$ to be an \emph{indeterminate} quantity that is defined only by its combining rules.

Finite multiples of $\epsilon$, such as $2\epsilon$ and $\epsilon/5$ are possible, and obey the obvious ordering.  $0<\epsilon/5 < \epsilon < 2\epsilon < r$.   These multiples are called \emph{first order infinitesimals}.   $\epsilon^2$ is a \emph{second order infinitesimal}.    $0<\epsilon^2<\epsilon$.   Finite multiples of $\epsilon^2$ operate similarly to first order infinitesimals.    Infinitesimals of any positive integral order are possible.  If $0<i<j$, then $\epsilon^j<\epsilon^i$.  Knuth has a charming novel on such numbers\cite{knuth-surreal}.

How is the predicate $(a<b)$ , where $a$ and $b$ are finite reals, affected by adding infinitesimals to $a$ and $b$?  The test might become $a+\epsilon^i<b+\epsilon^j$.  Assume, without loss of generality, that $0<i<j$.  If $a\ne b$, then if $a<b$ is true, then also $a+\epsilon^i<b+\epsilon^j$ is true.  Adding the infinitesimals didn't change the result.  However, if $a=b$, then the infinitesimals break the tie.  If $a=b$, then $a+\epsilon^i<b+\epsilon^j$ reduces to $\epsilon^i<\epsilon^j$, which reduces to $i>j$.

So, adding infinitesimals to $a$ and $b$ changes the predicate from  $(a<b)$ to
$(\mathit{if}\  (a\ne b) \ \mathit{then}\  (a<b)\  \mathit{else}\  (i>j))$ .
The execution time is increased by only the cost of one or two comparisons, which is probably insignificant in the context of the whole program.

\section{Simulation of Simplicity}

This section describes how to  break degeneracies with Simulation of Simplicity (SoS), \cite{Edelsbrunner:2002:TPS:2666864.2666893, Edelsbrunner:2001:SMI:378583.378644, edelsbrunner91, Levy:2016:REG:2884078.2884210, Schorn:1993:AAR:166614.166626}.  Note that the term \emph{simulation} is used in these computational geometry papers with a quite different meaning than used in the modeling and simulation community.

A degeneracy is, in a sense, a dimension reduction in the input, or a constraint between input parameters.   SoS has been used for computing contour trees \cite{Carr:2000:CCT:338219.338659}, triangulation \cite{Beichl:2002:DDT:766205.766281}, polyhedral modeling \cite{Fortune:1995:PME:218013.218065},  molecular modeling \cite{Halperin:1997:PSS:262839.262955}, computing line arrangements \cite{Chazelle:1991:CFA:127787.127864}, exact boundary evaluation \cite{Ouchi:2004:HDE:1217875.1217926},
and polygon overlay \cite{Audet-ParallelOverlay}.

These are the types of degeneracies that we wish to break (i.e., handle) with SoS: (i)  two different points having the same value, (ii) a point, possibly the endpoint of an edge, being incident on a line, or extended edge,  including the case of two edges being on the same infinite line, and (iii) two edges having the same slope.   The last degeneracy might not be an immediate problem, but is easy to handle.

SoS is a technique  to add infinitesimals of different orders to the points' coordinates.   The order cannot just be a function of a coordinate's value, but must depend on something unique to the point.   Therefore our SoS algorithm goes as follows.

\begin{enumerate}
\item Index all the coordinates of all the input points, from $0$ up.
\item Let the $i$-th coordinate be $x_i$.  So, point \#$k$ will have coordinates $(x_{2k}, x_{2k+1})$.
\item Modify the coordinates thus:  $x_i \rightarrow x_i+\epsilon^{2^i}$.
\end{enumerate}
This will break all degeneracies.   For instance, all possible edges between input points must have different slopes.
The above algorithm is general; however sometimes simpler SoS algorithms are adequate in special cases.


\section{Examples of SoS in use}

\subsection{Point on edge in 1D}\label{s:edge}

Testing whether a point is on an edge in 1D, while doing something useful when the point coincides with either endpoint, illustrates how this technique works.

\begin{wrapfigure}[5]{r}{.2\textwidth}
\includegraphics[width=.2\textwidth]{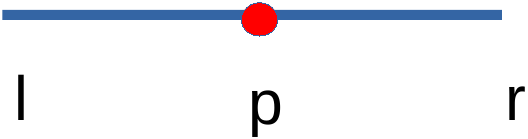}
  \caption{\small Point on edge}
  \label{f:point-on-edge}
\end{wrapfigure}

Consider the 1D case of a point with coordinate $p$ on an edge with left and right coordinates $l$ and $r$ respectively; see Figure \ref{f:point-on-edge}.   The point intersects the edge, in a set theoretic sense, if $(l\le p ) \wedge ( p\le r)$.  However, is this what we want  when the point is one of the endpoints?   We want an answer that makes this a useful subroutine of larger algorithms, such as point in polygon.   That is, we want to choose a definition of \emph{intersects} here that makes the larger algorithm \emph{correct}, for some useful definition of \emph{correct}, such as, not causing topological errors.

In this simple case, it is sufficient to modify $p$ thus: $p' \leftarrow p + \epsilon$ and use $p'$ instead of $p$.  The first version of the new predicate for testing point inclusion is now $(l\le p' ) \wedge ( p'\le r)$ .  However, if $l= p$ then $l<p'$, so $l\le p'$ is equivalent to $l\le p$.  Similarly if $p=r$ then $p'>r$ so $p'\le r$ is equivalent to $p<r$.

So the final SoS version of the point inclusion predicate is $(l\le p ) \wedge ( p< r)$.  Although we derived this using infinitesimals, it does not contain any infinitesimals.  It can be coded in any usual programming language using the usual arithmetic.  It also does not take a noticeable amount of extra time to execute.  The complexity associated with using SoS resides in the derivation of the new logic, not in the execution.  The resulting code will be obscure in that a viewer may not understand the underlying motivation.  That might be good or bad.

\subsection{Point in polygon test}\label{s:pnpoly}

Here we show how SoS solves the problem presented in the introduction on Page \pageref{f:polylines} of testing whether a query point  $p$ is inside a polygon $\mathcal P$.
For simplicity, do not consider the case of $p$ being on an edge of $\mathcal P$.  Also do not treat complicated types of $\mathcal P$, such as multiple
\begin{wrapfigure}[15]{r}{.3\textwidth}
\vspace{-.1in}
\includegraphics[width=.35\textwidth]{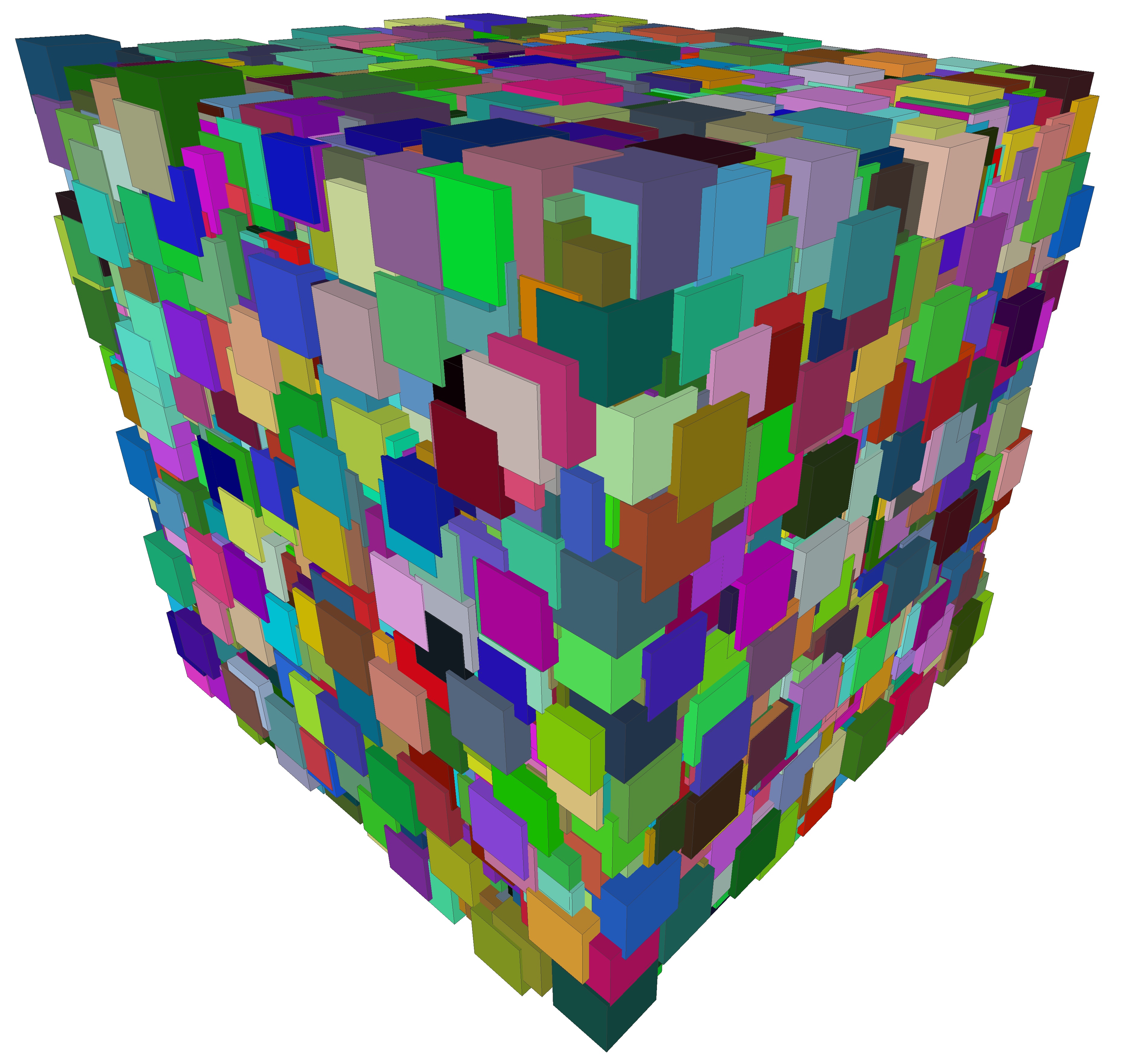}
  \caption{\small  Union of cubes}
  \label{f:union50}
\end{wrapfigure}
components and nested holes, although this algorithm handles them fine.


(There is another point testing algorithm that is more popular than good, which adds the angles subtended at $p$ by each edge of $\mathcal P$.  If the sum is $0$, $p$ is outside, if $2\pi$, then $p$ is inside.   The difficulty is that computing the subtended angle for an edge requires determining whether that edge crosses the positive x-axis, so we're back at the Jordan curve algorithm, but complexified with an arctan evaluation.)

Using the Jordan curve algorithm on polygon $ABCD$ in Figure \ref{f:pnpoly}, we want to count intersections of rays running up from query points with the edges of $ABCD$.  The ray from point $r$ has $0$ intersections; $r$ is outside.  The ray from $s$ has $1$; $s$ is inside.  $t$ induces $2$ intersections; it is outside.  What about the ray from $p$?  Since $p$ is outside, the total number of intersections with the edges $DA$ and $AB$ must be even; either $0$ or $2$.  However, also consider $q$, which is inside.  Its ray must have exactly one intersection with the two edges $AB$ and $BC$.  Which edge should that be?  In this case, the solution is to extend the algorithm in Section \ref{s:edge}.  The ray will intersect the edge, iff its x-coordinate passes that test and also the point is beneath the edge.

Here is the algorithm using SoS to test whether the ray from point $p=(p_x, p_y)$ intersects the edge of $\mathcal P$ with endpoints $v_i=(v_{ix}, v_{iy})$, $v_{i+1}=(v_{i+1,x}, v_{i+1,y})$.  $p$ is inside $\mathcal P$ iff the total number of intersections is even.

\begin{enumerate}
  \item If $v_{ix}=v_{i+1,x}$ then there is no intersection here, so return 0.
  \item (Identify the left and right ends of the edge.)  If $x_{ix}<v_{i+1,x}$ then let $l\leftarrow v_i$ and $r\leftarrow v_{i+1}$.   Else let $r
    \leftarrow v_i$ and $l\leftarrow v_{i+1}$.
\item 
  If  $(l_x > p_x ) \vee ( p_x \ge r_x )$ then there is no intersection here, so return 0.
\item 
Let $D\leftarrow
\begin{vmatrix} l_x & l_y & 1 \\ r_x & r_y & 1 \\ p_x & p_y & 1 \end{vmatrix} $.
\item If $D=0$ then $p$ is on the edge, which we are not considering, so return an error.
  \item Else, $p$ is below the edge iff $D<0$, so return 1 iff $D<0$ else return 0.

\end{enumerate}

\subsection{Volume of union of cubes}

This is another application of SoS to treat degeneracies.  The problem is to compute the volume, area, and edge length of the polyhedron resulting from the union of tens of millions of identical isothetic cubes.  The purpose is to test some fast, parallel, linear time, formulae for geometric mass properties.  The isothetic requirement removes any floating point roundoff issues.  Our algorithm \begin{wrapfigure}[13]{r}{.3\textwidth}
\includegraphics[width=.3\textwidth]{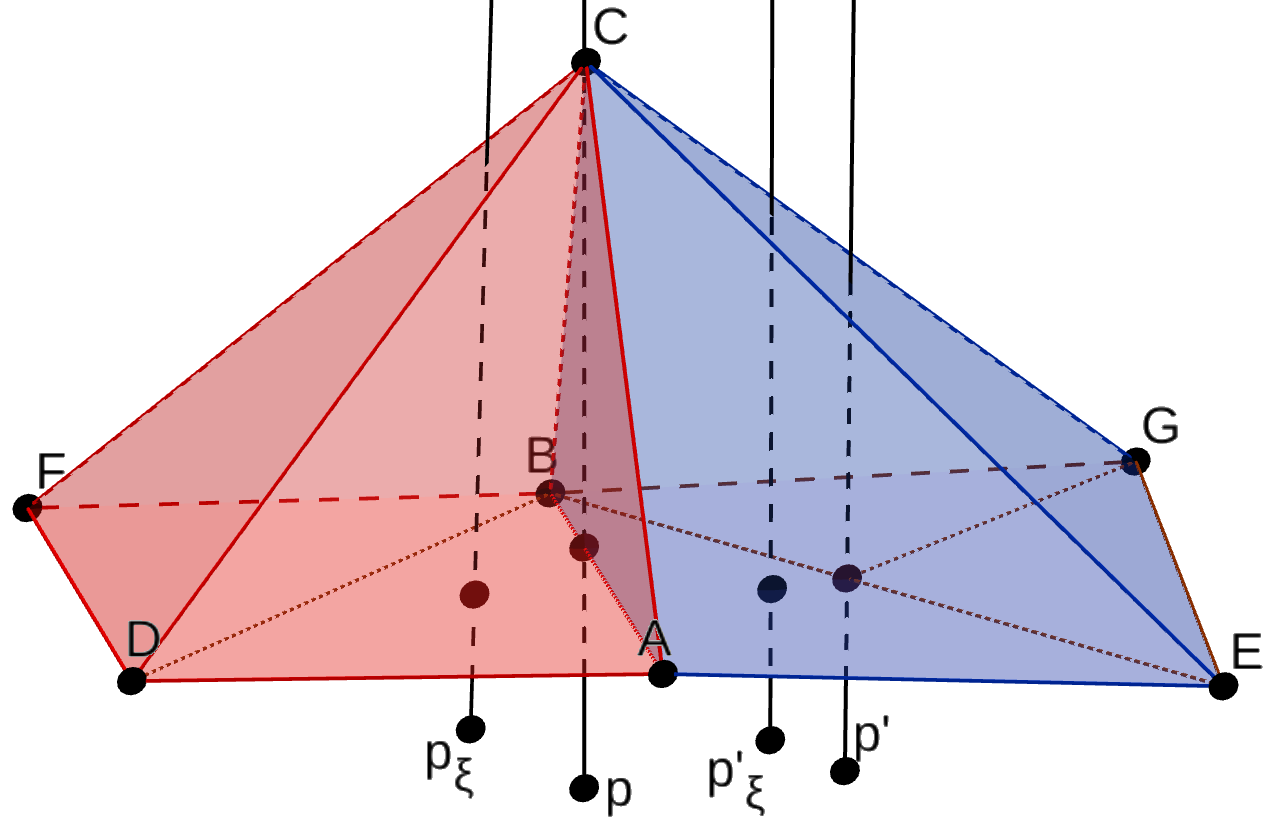}
  \caption{\small Ray casting through a 3D mesh, showing infinesimal perturbations}
  \label{f:pinmesh-perturb}
\end{wrapfigure}
starts by finding intersections of faces with edges, and intersections of three faces.  It also frequently tests whether a vertex is inside another cube.  Using several map-reduce operations it computes the desired mass properties.  Theoretical analysis and implementation on a multicore computer, with test results on up to 100M cubes, are presented in \cite{wrf-union-analysis-2004,wrf-dimacs2005,fwcg-2013}.  Figure \ref{f:union50} shows the union of a few cubes.  Note the complexity of the union, which is a difficult case for some algorithms.

A degeneracy occurs when a vertex of one cube is incident on a face of another cube.  This case subsumes all the other cases.   If these were not handled correctly, the output would be completely wrong.   This property helps to identify when the degeneracies are not being handled correctly.

SoS is used to treat the degeneracies.   The order of infinitesimals added is a function of the indexes of the edges and faces.   This is used to modify conditional tests of coordinates by adding tests on the indices of the edges and faces.   So, the execution time is not noticeably increased.
If the input is independently and identically distributed uniform random, then there is an equation for the expected output volume, and it agrees with what is computed.  This gives us confidence that SoS is working here.

\subsection{Point location in 3D mesh}
\begin{wrapfigure}[17]{r}{.3\textwidth}
\vspace{-.6in}

\includegraphics[width=.3\textwidth]{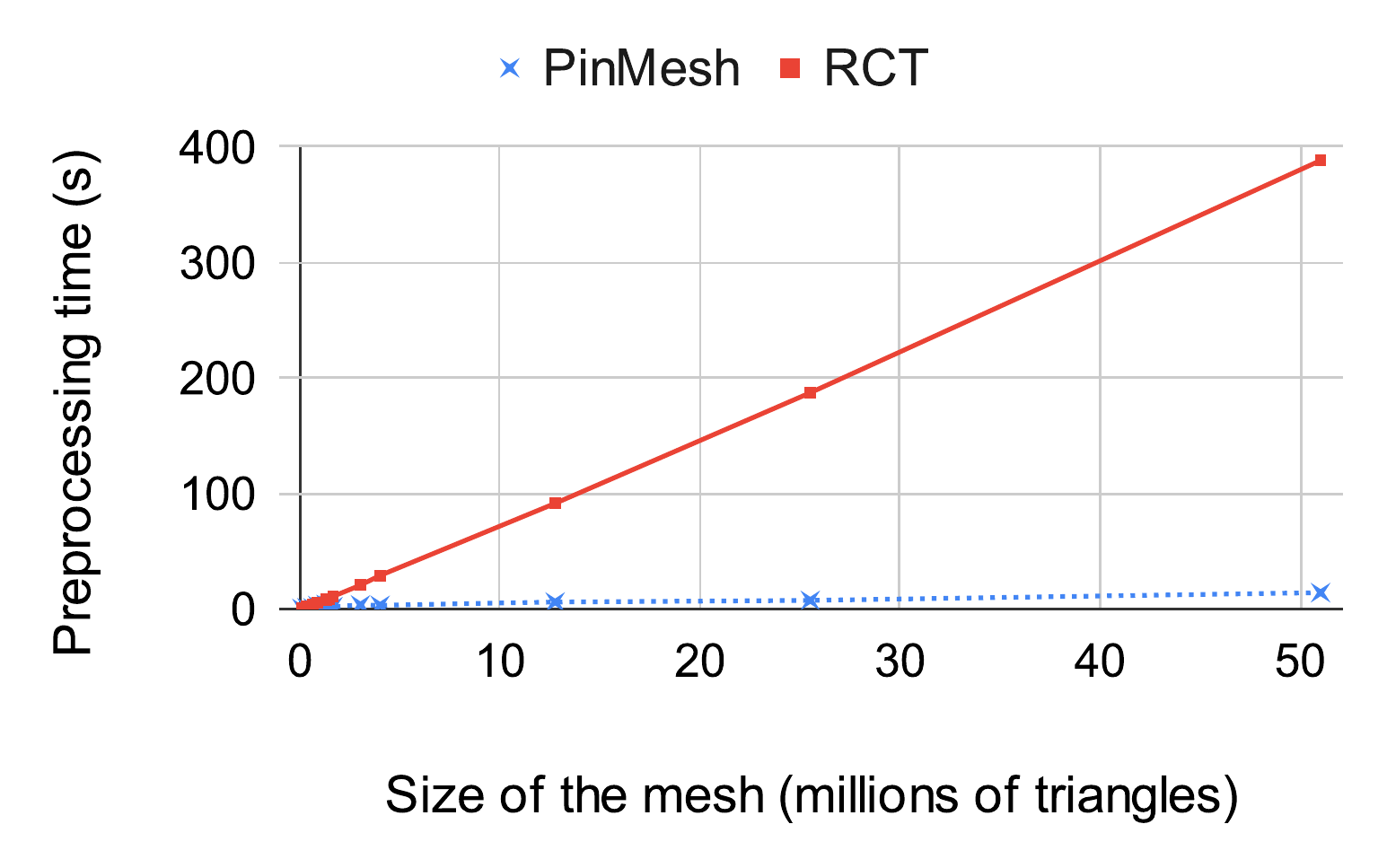}
\includegraphics[width=.3\textwidth]{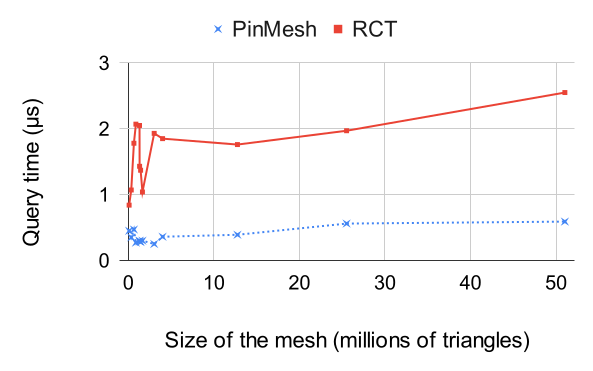}
  \caption{\small Point location preprocessing and query times on large 3D datasets compared to RCT}
  \label{f:pinmesh-times}
\end{wrapfigure}

This application, PinMesh, combines SoS with several other techniques to produce an implementation that is faster and more robust than other techniques.  This demonstrates that SoS works well with other techniques.
PinMesh preprocesses a polyhedral mesh, also known as a multimaterial mesh, to perform 3D point location queries, \cite{salles-pinmesh-smi-2016}. It combines several innovative components to efficiently handle the largest available meshes.  These include a 2-level uniform grid, exact arithmetic with rational numbers to prevent roundoff errors, and symbolic perturbation with Simulation of Simplicity (SoS) to handle geometric degeneracies or special cases. PinMesh is intended to be a subroutine in more complex algorithms.

Our implementation can preprocess a dataset and perform 1 million queries up to 27 times faster than RCT (Relative Closest Triangle), the current fastest algorithm.  Preprocessing a sample dataset with 50 million triangles took only 14 elapsed seconds on a 16-core Xeon processor. The mean query time was 0.6 microseconds.  In general, the preprocessing time was linear in the data size, while the query time was almost constant.   PinMesh also parallelized nicely on a multicore shared memory machine.  Figure \ref{f:pinmesh-perturb} illustrates infinesimallly perturbing a query point's location when casting a ray through a 3D mesh.    Figure \ref{f:pinmesh-times} presents comparative execution times  on large datasets.   Our implementation won a Reproducibility Stamp because the reviewers could reproduce our results.

\subsection{Intersecting 3D triangular meshes}
\begin{wrapfigure}[11]{r}{.3\textwidth}
\vspace{-.6in}
\includegraphics[width=.3\textwidth]{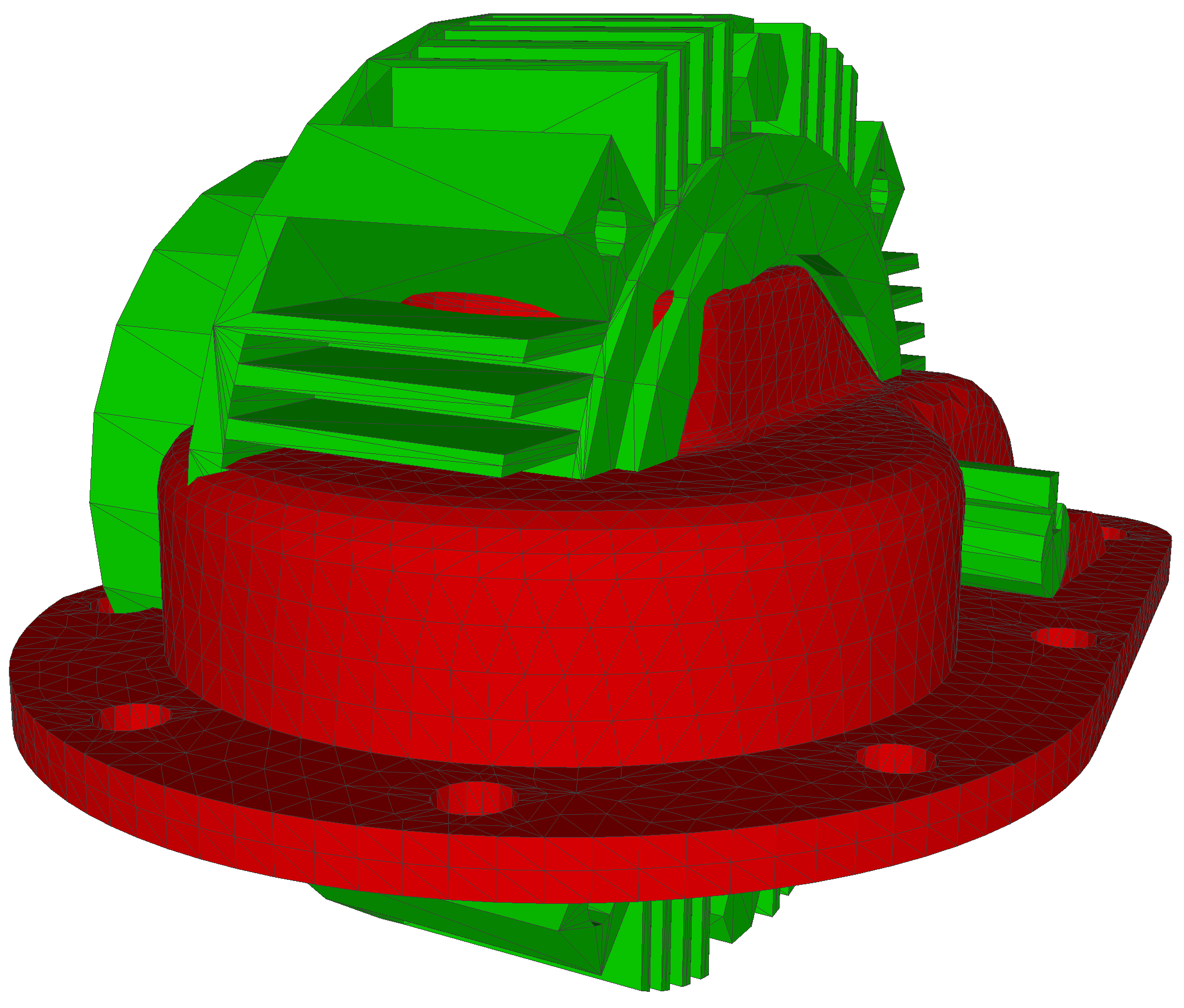}
  \caption{\small Two polyhedra being intersected}
  \label{f:salles-2019-04-cad-Fig4a}
\end{wrapfigure}

Our final application of SoS is 3D-EPUG-Overlay, a fast, exact, parallel, memory-efficient, algorithm for computing the intersection between two large 3-D triangular meshes with geometric degeneracies, \cite{salles-parallel-mesh-sigspatial-2017, salles-exact-cad-2019,marcelo-employing-gem-2022, marcelo-gpu-predicates-2021, mesh-orientation-s3pm-2017}; Figure \ref{f:salles-2019-04-cad-Fig4a}.  Handling degeneracies correctly is important because some other programs use heuristics with a user-supplied tolerance, and so sometimes fail.

Applications include CAD/CAM, CFD, GIS, and additive manufacturing.  3D-EPUG-Overlay combines 5 separate techniques: multiple precision rational numbers to eliminate roundoff errors during the computations; Simulation of Simplicity to properly handle geometric degeneracies; simple data representations and only local topological information to simplify the correct processing of the data and make the algorithm more parallelizable; a uniform grid to efficiently index the data, and accelerate testing pairs of triangles for intersection or locating points in the mesh; and parallel programming to exploit current hardware.  To simplify the symbolic perturbation, the algorithm employs only orientation predicates.

There is a challenge in the mesh intersection problem: the predicates will have not only to handle input vertices (with real or rational coordinates), but also vertices generated from intersections. Since the coordinates of a vertex generated from an intersection are a function of five input points (two points defining an edge of one mesh and three points defining a triangle of the other mesh) and these points are perturbed, then the orientation has to be modified to handle these points.  The 3D orientation will only be computed using, as arguments, three input vertices and another vertex that may be either an input vertex or a vertex from the intersection. Thus, at least two versions of the 3D orientation had to be implemented.


We successfully stress tested 3D-EPUG-Overlay by overlaying a polyhedron with translated or rotated versions of itself, in addition to overlaying pairs of different objects with up to 8M triangles.
   
\section{Summary and Acknowledgements}

Simulation of Simplicity, adding infinitesimals of different orders from a non-Archimedian ordered field to geometric coordinates, is a powerful technique to remove geometric degeneracies.  The infinitesimals are used to modify the program, which executes using the usual arithmetic.  The modified program often has the same length, or only a slight greater length, and has the same or only an insignificantly greater execution time.  The main limitation is that computations must be exact; SoS relies on exact equality tests.  The cost of this technique resides in the required analysis of the algorithm that this is being applied to.

Our code, albeit only research-quality, is generally freely available for nonprofit research and education.  We believe in our results and welcome stress tests and comparisons.

This research was partially supported by FAPEMIG, CAPES (Ciencia
sem Fronteiras - grant 9085/13-0), CNPq, and a gift from Dr Wenli Li.

\bibliographystyle{abbrv}

\small

\newcounter{hours}\newcounter{minutes}
\newcommand{\printtime}{%
\setcounter{hours}{\time/60}%
\setcounter{minutes}{\time-\value{hours}*60}%
\thehours:\theminutes}
\hfill \scriptsize\hspace{1.2in}\today, \printtime

\end{document}